\documentclass[aps,prd,reprint,nofootinbib,longbibliography,showkeys,dvipsnames]{revtex4-2}
\usepackage{ORCIDinREVTeX}

\usepackage{graphicx,wrapfig,float,slashed}
\usepackage{amsmath,amssymb,dsfont,epsfig,graphicx}
\usepackage{ragged2e}
\usepackage[utf8]{inputenc}
\usepackage{enumitem}
\usepackage{epstopdf}
\usepackage{xcolor}
\usepackage[toc]{appendix}
\usepackage{multirow}
\usepackage{hyperref}

\usepackage[normalem]{ulem}
\allowdisplaybreaks
\usepackage{pifont}

\usepackage{mathtools}

\newcommand{\nc}{\newcommand}
\nc{\non}{\nonumber}
\nc{\hc}{\hbox {h.c.}}
\nc{\noi}{\noindent}
\nc{\barx}{\bar{x}}
\nc{\pbarn}{\;\hbox {pb}}
\nc{\fbarn}{\;\hbox {fb}} 

\nc{\hsp}{\hspace{0.5cm}}
\nc{\lsp}{\hspace{1cm}}
\nc{\Lsp}{\hspace{2cm}}
\nc{\LLsp}{\lsp\lsp}
\nc{\lra}{\longrightarrow}
\nc{\p}{\prime}
\nc{\sgn}{\text{sgn}}
\nc{\ph}{\varphi}
\nc{\op}{{\cal O}}

\def\eq#1{Eq.~(\ref{#1})}
\def\fig#1{Fig.~\ref{#1}}
\def\sec#1{Sec.~\ref{#1}}
\def\tab#1{Table~\ref{#1}}

\nc{\beq}{\begin{equation}}  \nc{\eeq}{\end{equation}}
\nc{\bea}{\begin{eqnarray}}  \nc{\eea}{\end{eqnarray}}
\nc{\baa}{\begin{array}}     \nc{\eaa}{\end{array}}
\nc{\bit}{\begin{itemize}}   \nc{\eit}{\end{itemize}}
\nc{\ben}{\begin{enumerate}} \nc{\een}{\end{enumerate}}
\nc{\bce}{\begin{center}}    \nc{\ece}{\end{center}}
\nc{\bpm}{\begin{pmatrix}}   \nc{\epm}{\end{pmatrix}}
\nc{\bvt}{\begin{verbatim}}  \nc{\evt}{\end{verbatim}}

\def\lsim{\mathrel{\raise.3ex\hbox{$<$\kern-.75em\lower1ex\hbox{$\sim$}}}}
\def\gsim{\mathrel{\raise.3ex\hbox{$>$\kern-.75em\lower1ex\hbox{$\sim$}}}}

\def\udots{\mathinner{\mkern1mu\raise1pt\vbox{\kern7pt\hbox{.}}\mkern2mu\raise4pt\hbox{.}\mkern2mu\raise7pt\hbox{.}\mkern1mu}}

\def\gev{\;\hbox{GeV}}
\def\tev{\;\hbox{TeV}}

\definecolor{agray}{rgb}{0.95, 0.95, 0.99}

\def\G{{\cal G}}
\def\H{{\cal H}}

\nc{\cw}{\cos\theta_{W}}
\nc{\sw}{\sin\theta_{W}}
\nc{\cwsq}{\cos^2\theta_{W}}
\nc{\swsq}{\sin^2\!\theta_{W}}
\definecolor{cred}{rgb}{0.6, 0.0, 0.0}
\setlist[itemize]{itemsep=0em, topsep=0.3em}

\begin{document}
%%%%%%%%%%%%%%%%%%%%%%%%%%%%%%%%%%%%%%%%%%%%%%%%%%
\title{Conformal little Higgs models}

\author{Aqeel Ahmed}
\orcid{0000-0002-2907-2433}
\email{aqeel.ahmed@mpi-hd.mpg.de}
\affiliation{Max-Planck-Institut für Kernphysik, Saupfercheckweg 1, 69117 Heidelberg, Germany}
\author{Manfred Lindner}
\orcid{0000-0002-3704-6016}
\email{lindner@mpi-hd.mpg.de}
\affiliation{Max-Planck-Institut für Kernphysik, Saupfercheckweg 1, 69117 Heidelberg, Germany}
\author{Philipp Saake}
\orcid{0000-0002-5500-6827}
\email{saake@mpi-hd.mpg.de}
\affiliation{Max-Planck-Institut für Kernphysik, Saupfercheckweg 1, 69117 Heidelberg, Germany}

%%%%%%%%%%%%%%%%%%%%%%%%%
%\date{\today}

\begin{abstract}
Little Higgs models address the hierarchy problem by identifying the SM Higgs doublet as pseudo-Nambu--Goldstone bosons (pNGB) arising from global symmetries with collective breakings. These models are designed to address the little hierarchy problem up to a scale of $\Lambda\!\sim\! {\cal O}(10)$~TeV. Consequently, these models necessitate an ultraviolet (UV) completion above this scale. On the other hand, conformal extensions of the Standard Model are intriguing because scales emerge as a consequence of dimensional transmutation. In this study, we present a unified framework in which the electroweak hierarchy problem is tackled through a conformal symmetry collectively broken around the TeV scale, offering an appealing UV completion for little Higgs models. Notably, this framework automatically ensures the presence of the required UV fixed points, eliminating the need for careful adjustments to the particle content of the theory. Moreover, this framework naturally addresses the flavor puzzles associated with composite or little Higgs models. Furthermore, we suggest that in this framework all known little Higgs models can be UV-completed through conformal dynamics above the scale $\Lambda$ up to arbitrary high scales.
\end{abstract}


%%%%%%%%%%%%%%%%%%%%%%%%%
\maketitle
%\toccontinuoustrue

%%%%%%%%%%%%%%%%%%%%%%%%%
\section{Introduction}
\label{s.intro}
%%%%%%%%%%%%%%%%%%%%%%%%%

The Standard Model (SM) of particle physics works extremely well despite the fact that there are a number of theoretical and experimental reasons for embedding into more fundamental underlying theories. The so-called hierarchy problem is one of the main reasons to go beyond the SM. Composite Higgs models provide an attractive solution where the SM Higgs boson is a composite state of more elementary constituents with some strong confining dynamics~\cite{Kaplan:1983fs,Kaplan:1983sm,Georgi:1984af}. However such models naturally predict resonant states close to the electroweak scale and null results at the LHC for such states lead to the so-called little hierarchy problem. To alleviate this little hierarchy problem the little Higgs models were invented~\cite{Arkani-Hamed:2001nha,Arkani-Hamed:2002iiv,Arkani-Hamed:2002ikv,Low:2002ws,Kaplan:2003uc,Chang:2003un,Schmaltz:2004de,Schmaltz:2010ac}. The idea is that scalars are eventually not fundamental, but Goldstone bosons or pseudo-Nambu--Goldstone bosons (pNGBs) related to suitable larger symmetries and their collective breaking patterns. This implies that Higgs couplings and Yukawa couplings would not be fundamental, but effective such that only gauge couplings would survive in the UV limit.

Another interesting route to address the hierarchy problem is conformal symmetry. Theoretically it appears interesting since all scales would emerge dynamically instead of having relevant dimensionful operators in a lagrangian. This direction is also tempting from the fact that the SM\,\footnote{Without neutrino masses or with Dirac neutrino masses only.} is a one-scale theory where all masses appear from dimensionless couplings times one single scale, the electroweak vacuum expectation value (VEV), $v \simeq 174\gev$. Moreover, the independent parameters of the SM are such that the effective potential of the Higgs sector becomes miraculously flat at very high scales, see e.g.~\cite{Holthausen:2011aa}. This may point to a vacuum stability problem~\cite{Holthausen:2011aa,Degrassi:2012ry}, but being conservative about all theoretical and experimental errors it may also imply that a flat potential has a special meaning and is associated with massless particles like Goldstone bosons. Combined with a vanishing mass parameter this could indicate that conformal or shift symmetry plays a role at high scales. 

Implementing the idea of conformal symmetry one must avoid conformal anomalies since otherwise all benefits of the symmetry would be lost to quantum effects. This implies that all couplings of the theory must eventually have a UV fixed point, i.e. vanishing beta functions which are related to the trace of the energy-momentum tensor. Obtaining UV fixed points for all Higgs, Yukawa, and gauge couplings of a theory is a non-trivial requirement and the problem may be solved by carefully selecting the gauge groups and the representations~\cite{Houtz:2016jxk}. Even if one is successful the question remains if there is a principle behind the required specific choices. 

In this paper, we combine the ideas of composite Higgs models with conformal symmetry and implement them as a UV completion for the little Higgs models. Consequently, all scalar and Yukawa couplings are not fundamental such that only gauge couplings survive in the UV. This automatically guarantees that the theory has the required UV fixed points if it is based on non-Abelian gauge groups. In particular, we consider a UV theory with fundamental fermions, called {\it technifermions},  in a strongly coupled non-Abelian gauge theory such that we require that theory is in the conformal phase at the high energies with specific choices of fermion flavors and colors for the gauge group. The conformal symmetry is softly broken by relevant deformations of technifermion mass terms at conformal breaking scale~$\Lambda$. Since the theory is strongly coupled at energies close to the conformal breaking scale~$\Lambda$ it confines and breaks the residual chiral symmetry spontaneously such that it gives pNGBs related to the chiral symmetry breaking, see also \cite{Luty:2004ye,Luty:2008vs,Galloway:2010bp}. 
In this framework, the Higgs boson and other pNGBs emerge as composite states of the fundamental fermions. Furthermore, the symmetry structure in the model is such that the low-energy symmetry is broken collectively \`a\,la little Higgs models~\cite{Arkani-Hamed:2001nha,Arkani-Hamed:2002iiv,Arkani-Hamed:2002ikv,Low:2002ws,Kaplan:2003uc,Chang:2003un,Schmaltz:2004de,Schmaltz:2010ac}. Hence, this framework provides an interesting UV completion of the little Higgs models through conformal dynamics which makes these models consistent up to arbitrary high scales. 

As an explicit example, we consider the UV completion of the `bestest' little Higgs model~\cite{Schmaltz:2010ac} with strongly coupled conformal dynamics, based on global symmetry breaking in the coset $SU(4)^2/SU(4)$. We utilize four massless and four massive bifundamental technifermions, denoted as $\psi$ and $\chi$ respectively, all charged under the confining gauge group $SU(2)$. This choice places the gauge theory within the conformal window~\cite{Banks:1981nn,Appelquist:1998rb} above the scale~$\Lambda$. At this scale, the conformal symmetry is broken due to the relevant deformation caused by the massive technifermion bilinear operator. Below this scale, the gauge theory undergoes confinement and spontaneously breaks the flavor symmetry of four massless technifermions to $SU(4)^2/SU(4)$. Notably, the SM Higgs doublet is a pNGB that incorporates electroweak custodial symmetry.
We highlight that in conformal little Higgs models, the flavor scale, at which SM flavor physics, particularly the top-quark sector, becomes strongly coupled, can be naturally decoupled from the little Higgs breaking scale. For a related discussion regarding technicolor models, see~\cite{Luty:2004ye}. The model can in principle be distinguished from other UV completions of the little Higgs models by probes at scale $\Lambda$. We comment on the possible signatures of conformal little Higgs models at the future colliders probings energies $\op(10)\tev$. 
Moreover, we propose that any known little Higgs model based on symmetry breaking coset $\G/\H$ can be UV-completed with strongly coupled conformal dynamics by extending the fermionic content of the original flavor symmetry charged under a confining QCD-like gauge theory.

The paper is organized as follows: In the next \sec{s.CLH} we discuss the general framework of conformal little Higgs models based on flavor symmetry breaking $SU(N)^2/SU(N)$ of $N$ flavors charged under a confining gauge group $SU(N_c)$. We consider the case with $N=4$ and $N_c=3$ in \sec{s.model} which is the conformal UV completion of `bestest' little Higgs model. In \sec{s.general} we comment on the UV completion of any known little Higgs model with strongly coupled conformal dynamics and conclude in \sec{s.conc}. 

%%%%%%%%%
\section{Conformal UV completion of little Higgs models}
\label{s.CLH}
%%%%%%%%

In this section, we embed little Higgs models based on symmetry breaking coset $\G/\H$ into a strongly coupled conformal field theory (CFT), thus providing an attractive UV completion framework for little Higgs models valid up to arbitrary high scales. For concreteness, we focus on the case where $\G$ is a chiral global symmetry, $SU(N)_L\times SU(N)_R$, with $N$ Dirac fermion flavors spontaneously broken to its diagonal subgroup, $\H=SU(N)_V$, due to the strong confining dynamics of the gauge symmetry $SU(N_c)$. 
%This chiral symmetry breaking results in $(N^2-1)$ Goldstone bosons, among which is the SM Higgs doublet.

The key feature of our framework is that above the spontaneous symmetry-breaking scale, the theory exhibits conformal dynamics. To achieve this conformal dynamics, we add $N_m$ Dirac fermion flavors charged under an $SU(N_c)$ gauge symmetry, such that the theory has an enhanced global symmetry, $SU(N_f)_L\times SU(N_f)_R$, with $N_f\!\equiv\! N+N_m$ flavors. For non-supersymmetric gauge theories, it is argued that the strong conformal phase is achieved in the so-called conformal window with the $SU(N_c)$ gauge theory and $N_f$ flavors for~\cite{Banks:1981nn,Appelquist:1998rb}
\begin{equation}
\frac{7}{2}\lesssim \frac{N_f}{N_c}\lesssim\frac{11}{2}.		\label{CW}
\end{equation}
Since our interest lies in chiral symmetry breaking due to confining gauge dynamics for $N$ flavors, we consider these as massless Dirac fermions and denote them as $\psi_i$ ($i=1,\cdots,N$). Meanwhile, $N_m$ flavors are assumed to be massive Dirac fermions $\chi_j$ ($j=1,\cdots,N_m$) with mass~$M$. The technifermions $\psi_i$ and $\chi_j$ are in the fundamental representation of the gauge group $SU(N_c)$, whereas we also add $\hat\psi_i$ and $\hat\chi_j$ fermions in the conjugate (antifundamental) representation. We assume the Lagrangian for the UV theory contains
\beq
{\cal L}\supset {\cal L}_{\rm CFT}+ {\cal L}_{\rm def},
\eeq
where ${\cal L}_{\rm CFT}$ is the Lagrangian for the strongly coupled CFT and ${\cal L}_{\rm def}$ is the Lagrangian for a deformation to the CFT, which we consider of the following form, 
\beq
 {\cal L}_{\rm def}=-M\hat\chi\chi. 	\label{Ldef}
\eeq 
Above the mass parameter $M$ is a relevant deformation to the CFT for the scaling dimension of the $\hat\chi \chi$ bilinear operator $d\leq4$. Therefore, the above deformation breaks the conformal symmetry softly at the scale
\begin{equation}
\Lambda\equiv M^{1/(4-d)}.
\end{equation} 

To summarize, in this framework the strongly coupled CFT has an enhanced chiral symmetry \mbox{$SU(N_f)_L\!\times\! SU(N_f)_R$}. However, the theory exits the conformal fixed point below the scale $\Lambda$ due to deformation operator \eqref{Ldef} and enters the confining phase with $N$ flavors and $N_c$ colors\,\footnote{In general the conformal breaking scale $\Lambda$ can be different from the confining (condensation) scale of the QCD-like theory. However we assume that the CFT is strongly coupled, therefore the two scales can be identified as the same scale $\Lambda$.}. This confinement forms a condensate 
\beq
\langle\hat\psi_i^\alpha\psi_{j\alpha}\rangle\sim\frac{\Lambda^d}{(4\pi)^2}\, \delta_{ij},
\eeq 
which spontaneously breaks the flavor symmetry \mbox{$SU(N)_L\!\times\! SU(N)_R/SU(N)_V$}, where $\alpha$ is the $SU(N_c)$ color index and $i,j$ label the fermion flavors, i.e. $i,j\!=\!1,\cdots,N$. This symmetry breaking gives $(N^2-1)$ Goldstone bosons, $\hat\psi\psi$, corresponding to broken generators. Note the fermion bilinear scalar operator $\hat\psi \psi$ has the same scaling dimension $d$ as that of $\hat\chi\chi$ operator. This is due to the fact that above the conformal symmetry breaking, they belong to the same symmetry group, $SU(N_f)_L\!\times\! SU(N_f)_R$, which exhibits conformal dynamics. Due to unitarity considerations, the scaling dimension $d$ of these scalar operators $\hat\psi \psi$ (and $\hat\chi\chi$) is bounded from below to be greater than one, i.e. $d\geq 1$. The limiting case $d=1$ implies free (elementary) scalar fields. Whereas, $d=3$ for technicolor models~\cite{Susskind:1978ms,Weinberg:1975gm} and $d\gtrsim2$ can be achieved in walking technicolor models~\cite{Holdom:1984sk,Akiba:1985rr,Yamawaki:1985zg,Appelquist:1986an,Appelquist:1987fc,Appelquist:1986tr}. As we discuss below for a conformal little Higgs model we would require $1<d<2$. 

\begin{figure}[t!]
\centering{
\includegraphics[scale=0.9]{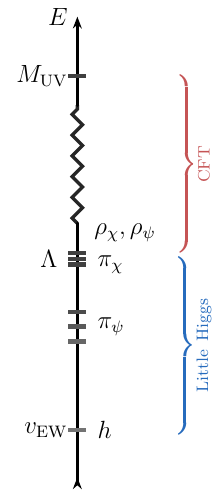}
}
\caption{A schematic depiction of energy scales in the conformal little Higgs model. In the UV we expect the theory enters a conformal fix point at scale $M_{\rm UV}$ a'la Banks-Zaks~\cite{Banks:1981nn}. At an IR scale $\Lambda$, the conformal symmetry is softly broken by a relevant deformation due to explicit mass parameter $M=\Lambda^{4-d}$ of the technifermion bilinear operator $\hat\chi\chi$. Below the conformal breaking scale $\Lambda$ the residual chiral symmetry with $N$ flavors is spontaneously broken by the confinement resulting in ($N^2-1$) pNGB states $\pi_\psi=\hat\psi\psi$ including the SM model Higgs boson $h$.
}\label{fig_scales}
\end{figure}

Below the conformal breaking scale $\Lambda$, the low-energy effective theory exhibits the little Higgs dynamics for the coset $\G/\H$ through the collective breaking patterns. This ensures that in the low-energy theory, the SM Higgs field and other pNGB masses are protected from quadratic divergences. Therefore, low-energy observables and predictions remain similar to those of any conventional little Higgs model. However, there are crucial differences between our conformal little Higgs model and other little Higgs UV completions that appear at the conformal breaking scale $\Lambda$. For instance, a general problem with strongly coupled UV completions of little Higgs or composite Higgs models is that flavor dynamics does not decouple from the electroweak breaking dynamics. Whereas, in the strongly coupled conformal little Higgs models, the flavor problem can be decoupled from the electroweak scale up to flavor scale $\Lambda_t$~\cite{Luty:2004ye},
\beq
\Lambda_t\equiv \Lambda \bigg(\frac{4\pi v}{m_t}\bigg)^{1/\epsilon}, \quad {\rm for} \quad \epsilon\equiv d-1\sim 1/{\rm few},		\label{Lam_t}
\eeq 
where $m_t$ is the top-quark mass and $v$ is the electroweak VEV. Furthermore, at energies at or above the conformal breaking scale $\Lambda$, the observable effects of strong conformal dynamics begin to appear in the form of broad resonances corresponding to $\rho_\psi$-mesons (corresponding to excited spin-1 $\hat\psi\psi$ states) and other composite states of the conformal dynamics (e.g., $\pi_\chi\equiv \hat\chi\chi$ states). In \fig{fig_scales} a schematic depiction with different energy scales is presented for our strongly coupled conformal little Higgs model. 

Note that in this framework, the Higgs operator $\op_H \sim \hat \psi\psi$ has a scaling dimension $d$, which needs to be close to 1 to address the flavor puzzle. On the other hand, for $d\to 1$, the Higgs operator becomes weakly coupled, and the scaling dimension $\Delta$ of the lowest gauge-invariant operator $\op_H^\dag\op_H$ becomes close to $2d$, i.e., $\Delta\to 2d$ for $d\to 1$. Therefore, the SM hierarchy problem reemerges. To avoid this problem, one requires that the scaling dimension of the $\op_H^\dag \op_H$ operator be irrelevant, i.e., $\Delta>4$~\cite{Luty:2004ye}, which can be achievable in a strongly coupled CFT. For a generic CFT theory, the limit $d\to 1$ was calculated in Ref.~\cite{Rattazzi:2008pe,Rychkov:2009ij}, where they obtained a general result:
\beq
\Delta \lesssim 2d + \op(\sqrt{d-1}),
\eeq
which implies roughly $d\gtrsim 1.5$ to have $\Delta\gtrsim 4$. Strictly speaking, the results of Ref.~\cite{Rattazzi:2008pe} cannot be directly applied to our case as they do not distinguish between the scalar operators that differ only by internal symmetries. Therefore, to address the (large) hierarchy problem, we require that $d$ is large enough to render the gauge-invariant scalar operator $\op_H^\dag\op_H$ irrelevant and small enough to effectively decouple the flavor puzzle\eqref{Lam_t}. 

%%%%%%%%%
\section{An example class of models}
\label{s.model}
%%%%%%%%

In this section, we consider an example class of little Higgs models with a global symmetry-breaking coset $SU(4)_L \times SU(4)_R / SU(4)_V$ with four light flavors of fundamental fermions. This example is based on the isomorphic coset of the bestest little Higgs model~\cite{Schmaltz:2010ac}. However, as we comment below, under certain low-energy assumptions, this example class can also lead to the `minimal moose' little Higgs model with custodial symmetry~\cite{Chang:2003un}, which is based on the coset $SO(5) \times SO(5) / SO(5)$. The aim of this section is to provide one example class of models with conformal UV dynamics that leads to little Higgs models below the conformal breaking scale $\Lambda\sim O(10)\tev$. We argue in the following that such UV completion of little Higgs models can lead to a class of little Higgs models with different low-energy properties based on symmetry-breaking patterns.

As mentioned above one of the main aspects of this work is to consider the UV completion of little Higgs models without introducing any elementary/fundamental scalars in the model.
For this purpose, we employ a confining gauge symmetry $SU(N_c)$ where for concreteness we take $N_c=3$ and new fermions $\psi$'s and $\chi$'s, labeled as {\it technifermions}, charged under this gauge group. In \tab{tab.techi_fermions} we summarize the quantum numbers for the new fermions under the confining gauge group as well as under the SM gauge group $SU(3)_{C}\times SU(2)_{L}\times U(1)_{Y}$. Where $\tilde\psi$ and $\psi^\prime$ are label as fundamentals of the electroweak gauge group $SU(2)_L\subset SU(4)_L$ and the custodial group $SU(2)_{L}^\prime\subset SU(4)_L$, respectively. The $U(1)_Y$ charges are associated with the diagonal generators of the custodial $SU(2)$. Similarly the conjugate fields $\hat \psi$ are charged under the subgroups of $SU(4)_R$.
\begin{table}[t!]
\centering{
\begin{tabular}
{|c|c|c|c|c|c}
\hline
& $SU(N_c)$ & $SU(3)_{C}$ & $SU(2)_{L}$ & $U(1)_{Y}$  \\
\hline \hline 
$\tilde\psi \equiv\bpm \psi_1\\ \psi_2 \epm $ & $\square $& $\mathbf{1} $& $\square $& 0 \\
$\psi^\prime \equiv\bpm \psi_3\\ \psi_4 \epm $ & $\square $& $\mathbf{1} $& $\begin{array}{c}
\mathbf{1} \\ \mathbf{1}
\end{array}$& $\begin{array}{c}
-\frac12 \\ +\frac12
\end{array}$\\
$\chi\times N_m$ & $\square $& $\mathbf{1} $& $\mathbf{1}$& 0\\
\hline 
\end{tabular}}
\caption{Quantum numbers for technifermions under the $SU(N_c)$ confining gauge group and the SM gauge group $SU(3)_{C}\times SU(2)_{L}\times U(1)_{Y}$.}
\label{tab.techi_fermions}
\end{table}

We assume the number of flavors $N_m$ of $\chi$ are such that the theory is near a strongly coupled conformal fixed point. 
In the following, we consider $N_m=8$ which is believed to be enough flavors to make the confining $SU(3)$ gauge theory strongly coupled nearly conformal fixed point. 
The Lagrangian for the confining sector is 
\beq
{\cal L}\supset -M_\psi \hat{\psi}\psi -M \hat{\chi} \chi,
\eeq
where $\psi=(\tilde\psi,\psi^\prime)$.
Note that the coefficients of ``mass'' terms, i.e. fermion bilinears, denoted with $M_{\psi}$ and $M$ has scaling dimension different from the canonical mass dimension~1. For a non-trivial scaling dimension $d$ of the fermion bilinears, which is a scalar operator, we have $4-d$ scaling dimension for $M_{\psi}$ and $M$. As mentioned above, for the scaling dimension $1<d<4$ these mass terms are relevant deformations that take the theory out of the conformal fixed point. In other words, the conformal symmetry is broken at scale $\Lambda$ where these deformations become strong. Since we are interested in $SU(4)_L\times SU(4)_R$ chiral breaking we assume $M_{\psi}=0$ such that the conformal breaking is dominated by $M$ deformation of $\bar \chi\chi$ operator and $\psi$'s are the massless flavor which breaks the chiral symmetry only due to the condensate $\langle\hat{\psi}\psi\rangle$. 

Since the technifermions $\chi$'s are heavy with masses of the order conformal breaking scale, one can integrate them out at the scale $\Lambda$. In the effective theory at/below scale $\Lambda=M^{1/(4-d)}$ we are left with four flavors of fermions $\psi$ and their conjugate fermions $\hat \psi$, which has $SU(4)_L\times SU(4)_R$ chiral symmetry. 
The technifermions $\psi$ and $\hat \psi$ transform as ${\bf 4}$ and $\bar {\bf 4}$ of $SU(4)_L$ and $SU(4)_R$, respectively, therefore, the condensate transforms as
\beq
\langle\hat \psi \psi \rangle=({\bf 4}, \bar {\bf 4})_{SU(4)_L \times SU(4)_R}\,.
\eeq
The above condensate breaks flavor symmetry to its diagonal subgroup, i.e. \mbox{$SU(4)_L \times SU(4)_R \to SU(4)_V$}, resulting in $15$ Goldstone bosons which transform as the adjoint of the unbroken diagonal $SU(4)_V$. 
Assuming massless technifermions~$\psi$ we can align the vacuum along the direction that preserves $SU(4)_V$ and does not break the SM EW symmetry,
\beq
\langle\hat\psi \psi \rangle=\frac{\Lambda^d}{(4\pi)^2} \bpm
\mathds{1} && 0 \\
0 && \mathds{1}
\epm,
\eeq
where $\Lambda$ is the scale of conformal symmetry breaking and $d$ is the scaling dimension of the condensate.

Before moving forward, we would like to highlight a crucial point. Starting with the $SU(4)^2$ flavor symmetry, one can also achieve a conformal UV completion for the `minimal moose' little Higgs model with custodial symmetry. This model is based on the coset $SO(5)^2/SO(5)$~\cite{Chang:2003un}, which is isomorphic to $Sp(4)^2/Sp(4)$. The procedure involves explicitly breaking the $SU(4)^2$ flavor symmetry down to $Sp(4)^2$ by introducing the following non-renormalizable term in the Lagrangian:
\beq
\mathcal{L}\supset \frac{c\, m^2}{\Lambda^{2d-2}}\mathrm{Tr}\Big[(\hat\psi \psi)J (\hat\psi \psi)^T J\Big],
\eeq
where $c\sim (4\pi)^2$ represents the strong coupling, $m\sim \Lambda$ is the mass scale, and $J$ is a matrix that preserves the above interaction under only $Sp(4)$ rotations while explicitly breaking the remaining orthogonal directions of $SU(4)$:
\beq
J\equiv \frac{1}{2}\bpm i\sigma^2 & 0 \\ 0 & i\sigma^2 \epm, \qquad \sigma^2= \bpm 0 & -i \\ i & 0 \epm.
\eeq
As a result, 5 of the 15 would-be Goldstone bosons acquire a mass of order $m\sim \Lambda$. Subsequently, the remaining $Sp(4)_L\times Sp(4)_R$ symmetry is broken by confining strong dynamics to the coset $Sp(4)^2/Sp(4)$, resulting in 10 Goldstone bosons, as discussed in~\cite{Batra:2007iz}. The complete symmetry breaking pattern and low-energy properties, such as the quartic Higgs coupling, of the little Higgs model of Ref.~\cite{Chang:2003un}, can be obtained by multiple repetitions of the above-described $Sp(4)^2/Sp(4)$ symmetry breaking.

In the following, we focus on the case with a full $SU(4)_L\times SU(4)_R$ symmetric case of the little Higgs model and highlight some of its properties.

%%%%%%%%%%%%%%%%%%
\subsection{Higgs sector}
\label{s.higgs}
%%%%%%%%%%%%%%%%%%%%

The chiral symmetry breaking of the coset $SU(4)_L\times SU(4)_R/SU(4)_V$ leads to $15$ Goldstone bosons which transform under the custodial symmetry $SO(4)\simeq SU(2)_L \times SU(2)_R\subset SU(4)_V$ as
\beq
\mathbf{15}_{SU(4)_{V}}=(2,2)+(2,2)+(3,1)+(1,3)+(1,1),
\eeq
where we have two doublets, two triplets, and a singlet under the custodial symmetry. We can write the Goldstone matrix as, 
\beq
U=e^{i2\Pi/f};
\eeq
where
\beq
\Pi=\frac{1}{2}\bpm
\sigma^a \Delta^a_1+\eta / \sqrt{2} & -i \Phi_H \\
i \Phi_H^{\dagger} & \sigma^a \Delta_2^a-\eta/ \sqrt{2}
\epm. 
\label{eq.GoldstoneMatrix}
\eeq
Above $\Phi_H$ is a bi-doublet,
\beq
\Phi_H \equiv \left(\widetilde H_1+i \widetilde H_2, \quad H_1+i H_2\right),
\eeq
with $\widetilde H_i\equiv i \sigma_2H^\ast_i$ where $H_i$ are Higgs doublets, whereas the triplets $\sigma^a \Delta^a_{1,2}$ are, 
\beq
\sigma^a \Delta^a=\left(\begin{array}{cc}
\Delta^0 & \sqrt{2} \Delta^{+} \\
\sqrt{2} \Delta^{-} & -\Delta^0
\end{array}\right).
\eeq
Note that the matrix $U$ transforms linearly under the $SU(4)_V$ group as, $U\to \Omega_V\cdot U \cdot \Omega_V^\dag$.

The EW symmetry breaking vacuum is where both the Higgs doublets acquire non-zero VEVs, i.e. $\langle H_{1,2}\rangle\equiv v_{1,2}$. Such that most general custodial symmetry preserving vacuum can be written as, 
\beq
\langle \Phi_H\rangle = v\,e^{i\beta}\mathds{1}, 
\eeq
where $v\! \equiv\! \sqrt{v_1^2+v_2^2}$ is the electroweak VEV and \mbox{$\tan\beta\equiv v_2/v_1$}. The misalignment of the EW vacuum $v$ with respect to the global symmetry breaking vacuum $f\equiv \Lambda/(4\pi)$ can be parametrized by a transformation, 
\beq
\Sigma= \Omega_0\cdot U\cdot \Omega_0,
\eeq
such that the new vacuum is, 
\beq
\langle \Sigma\rangle\equiv \Sigma_0 = \Omega_0\cdot \Omega_0\equiv \bpm
\cos \theta\, \mathds{1} & e^{i \beta} \sin \theta\, \mathds{1} \\
-e^{-i \beta} \sin \theta \, \mathds{1} & \cos\theta\, \mathds{1}
\epm,	
\eeq
where the vacuum rotation matrix is
\beq
\Omega_0= \bpm
\cos \frac{\theta}{2}\, \mathds{1} & e^{i \beta} \sin \frac{\theta}{2}\, \mathds{1} \\
-e^{-i \beta} \sin \frac{\theta}{2} \, \mathds{1} & \cos \frac{\theta}{2}\, \mathds{1}
\epm,
\eeq
with the misalignment angle $\sin\theta\equiv v/f$. In this EW symmetry breaking vacuum, the chiral condensate $(\hat \psi \psi)$ is related to $\Sigma$ as, 
\beq
(\hat \psi \psi) = \frac{\Lambda^d}{(4\pi)^2}\Sigma,
\eeq
where the factor of $(4\pi)^2$ is the Naïve Dimensional Analysis (NDA) value in a strongly coupled theory. 

One of the crucial aspects of the Little Higgs framework is the collective contribution to the Higgs quartic potential without generating the corresponding quadratic term. This feature can be introduced in our example model similarly to the `bestest Little Higgs' model~\cite{Schmaltz:2010ac}. For this purpose, we introduce two projection operators, 
\beq
J_1=\frac{1}{2}\bpm i\sigma^2 & 0 \\ 0 & i\sigma^2 \epm, \qquad J_2=\frac{1}{2}\bpm i\sigma^2 & 0 \\ 0 & -i\sigma^2 \epm,
\eeq  
such that the collective quartic potential is obtained as, 
\begin{align}
{\cal L}&\supset -\frac{\lambda_{12}}{4} f^4 \Big\vert\mathrm{Tr}\big(\Sigma\, J_1 \,\Sigma^T J_2\big)\Big\vert^2-\frac{\lambda_{21}}{4} f^4 \Big\vert\mathrm{Tr}\big(\Sigma\, J_2 \,\Sigma^T J_1\big)\Big\vert^2,		\label{eq.coll_pot}
\end{align}
where $\lambda_{ij}$ are dimensionless parameters. Note in the above Lagrangian, the first term breaks $SU(4)_L\times SU(4)_R$ symmetry to $Sp(4)_{L1}\times Sp(4)_{R2}$, whereas the second term breaks $SU(4)_L\times SU(4)_R$ symmetry to $Sp(4)_{L2}\times Sp(4)_{R1}$. Here $Sp(4)_{Li}$ (or $Sp(4)_{Ri}$) refers to two different vacuum configurations corresponding to $J_i$, with $i=1,2$. Such that the above terms only generate potential for the $\eta$ field, whereas all other pNGBs are protected by symmetries. In particular, $Sp(4)_{L1}\times Sp(4)_{R2}$ collectively protect the two Higgs doublets $H_1$ and $H_2$. For instance, in the electroweak symmetric phase, i.e., $\Sigma=U$, we can rewrite the above potential terms as,
\begin{align}
{\cal L}&=-\frac{\lambda_{12}}{2} f^2\Big(\eta + \frac{1}{\sqrt{2}f}(H_1^\dag H_2+\hc) +\cdots\Big)^2 \notag\\
&\quad -\frac{\lambda_{21}}{2} f^2\Big(\eta - \frac{1}{\sqrt{2}f}(H_1^\dag H_2+\hc)+\cdots \Big)^2,
\end{align}
where ellipses denote higher-order terms. The above Lagrangian generates only mass for the $\eta$ field,  
\beq
m_\eta^2=(\lambda_{12}+\lambda_{21})f^2,		\label{eq.m_eta}
\eeq
which can be integrated out at the tree level, generating a quartic term for the Higgs potential of the form, 
\beq
{\cal L}\supset-\frac{\lambda_{12}\lambda_{21}}{\lambda_{12}+\lambda_{21}}\Big(H_1^\dag H_2+\hc \Big)^2.
\eeq
Note the above potential has the desired form of the Little Higgs models employing the collective breaking involving two different couplings. 
We note that the above collective potential terms \eqref{eq.coll_pot} lead to the generation of the Higgs mass at one-loop order of the following form:
\beq
\frac{\lambda_{12}\lambda_{21}}{16\pi^2}f^2\,\ln\!\bigg(\frac{\Lambda^2}{\mu^2}\bigg)\Big(H_1^\dag H_1+H_2^\dag H_2\Big),
\eeq
which is only logarithmically sensitive to the cutoff $\Lambda$. Here, the renormalization scale $\mu$ is of the order $m_\eta$~\eqref{eq.m_eta} to minimize the finite corrections to the potential.
We refer to Ref.~\cite{Schmaltz:2010ac} for further details on the low-energy phenomenology of the Higgs sector as one can follow similar steps for this model.

%%%%%%%%%%%%%%%%%%%%
\subsection{Gauge sector}
\label{s.collective}
\begin{figure}[t]
\centering{
\includegraphics{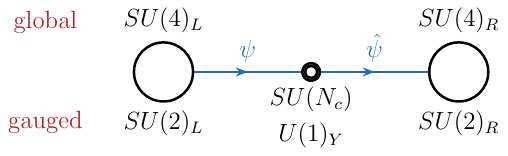}
}
\caption{Collective symmetry breaking the structure of the `bestest' conformal little Higgs.}\label{fig.gauge}
\end{figure}

%%%%%%%%%%%%%%%%%%%%
To employ the collective breaking in the gauge sector, we gauge subgroups $SU(2)_L\subset SO(4)_L\subset SU(4)_L$ and $SU(2)_R\subset SO(4)_R\subset SU(4)_R$. The symmetry structure of the theory can be depicted as a Moose diagram~\fig{fig.gauge}.
The gauge boson masses can be obtained as,
\begin{align}
{\cal L}&\supset \frac{1}{2}f^2{\rm Tr}\Big[(D_\mu\Sigma)^\dag(D^\mu\Sigma)\Big],		\label{Lgauge}\\
&\supset \frac{1}{2}g_{\rm EW}^2 f^2 \sin^2\!\theta W_\mu W^\mu+\frac{1}{4\cos^2\theta_W}g_{\rm EW}^2 f^2 \sin^2\!\theta Z_\mu Z^\mu 		\notag\\
&\quad+\frac{1}{2}(g_L^2+g_R^2) f^2 \cos^2\theta W_\mu^{\prime} W^{\prime \mu}\notag\\
&\quad +\frac{1}{4}(g_L^2+g_R^2) f^2 \cos^2\theta Z_\mu^\prime Z^{\prime\mu}+ \cdots. 
\end{align}
Above, ellipses denote Goldstone bosons' interactions with the gauge bosons. 
The covariant derivate is defined as, 
\beq
D_\mu \Sigma\equiv \partial_\mu \Sigma + i g_L {\cal A}_L^a T^a\Sigma - i g_R {\cal A}_R^a T^a\Sigma,
\eeq
where ${\cal A}_{A,B}$ are the gauge bosons corresponding to $SU(2)_{LA,B}$ and $g_{\rm EW}^2=g_L^2g_R^2/(g_L^2+g_R^2)$. 
Above we neglect the $U(1)_Y$ gauge bosons as its relevance to the little hierarchy problem is less severe. It is straightforward to include.
Masses of the heavy gauge bosons $W^\prime$ and $Z^\prime$ are, 
\beq
m^2_{W^\prime}=m^2_{Z^\prime}=\frac{1}{2}(g_L^2+g_R^2) f^2 \cos^2\theta, 	\label{e.mWp}
\eeq
whereas the SM gauge bosons are, 
\beq
m^2_{W}=m^2_{Z}\cos^2\theta_W =\frac{1}{2}g_{\rm EW}^2 v^2+\op(v^4/f^2), 
\eeq
where $\theta_W$ is the Weinberg angle.

One of the key features of the little Higgs models is that the Higgs boson and other uneaten pNGB masses are not quadratically sensitive to the symmetry-breaking scale $\Lambda$. 
It is straightforward to verify this by analysing the radiative corrections to the pNGB masses.
At one loop level, the radiative corrections to the Higgs and other uneaten pNGB masses due to light and heavy gauge bosons are, 
\beq
{\cal L}\!\supset\!  \frac{9 g_{\mathrm{EW}}^2 m_{W^{\prime}}^2}{64 \pi^2} \ln\!\bigg(\!\frac{\Lambda^2}{m_{W^{\prime}}^2}\!\bigg)\bigg[|H_1|^2+|H_2|^2+\frac{8}{3} |\Delta_1|^2\bigg],
\eeq
which is only sensitive to the heavy gauge boson mass which could be much smaller than the scale $\Lambda$. 

Note that in a strongly coupled theory, the dynamics that result in the nonlinear sigma model for the gauge sector \eqref{Lgauge} with weak gauge bosons of the theory would become strong at the scale $m_\rho$. In our framework, this scale is given by
\beq
m_\rho = \Lambda \bigg(\frac{4\pi v}{m_W}\bigg)^{1/(d-1)}\,,
\eeq
which can be naturally decoupled from the strong dynamics responsible for the electroweak symmetry breaking for \mbox{$\epsilon\equiv d-1\sim 1/{\rm few}$}.
For $\epsilon =1/3$ the gauge bosons strong dynamics scale $m_\rho$ can be pushed up to $\op(10^4)\tev$ for $\Lambda\sim\op(1)\tev$.

Here, we would like to note that this work aims to emphasize the main features of the strongly coupled conformal UV completion of Little Higgs models, while some of the low-energy phenomenological aspects are model-dependent and can be realized within this framework. For example, in the original `bestest Little Higgs' model~\cite{Schmaltz:2010ac}, it was argued that it is more favorable phenomenologically to have gauge partners heavier than the symmetry-breaking scale $f$. To achieve this, a modular approach was taken, involving two independent sigma models where a new scalar field was introduced to break a new global $SU(2) \times SU(2)$ symmetry to the diagonal $SU(2)$ at a higher scale $F>f$. It is worth noting that at or below the conformal breaking scale, a similar modular approach can naturally be adopted, where a new global symmetry $SU(2) \times SU(2)$ spontaneously breaks to the diagonal $SU(2)$ and is orthogonal to our strong sector and is weakly gauged. 
One possibility could be that two vector-like fermions, denoted as $\chi^\prime_{1,2}$, are charged under a new strong confining gauge group $SU(N_c)^\prime$ (orthogonal to the $SU(N_c)$ of the conformal sector) and are also charged under the new global symmetry $SU(2)_L^{\prime} \times SU(2)_R^{\prime}$. This setup leads to the confinement of $SU(N_c)^\prime$ and the condensation of $\langle \hat\chi^{\prime}\chi^\prime\rangle$, resulting in the breaking of the global symmetry $SU(2)_L^{\prime} \times SU(2)_R^{\prime}$ to orthogonal $SU(2)_V^\prime$ at the breaking scale $F\lesssim \Lambda$.
This approach naturally increases the masses of gauge partners~\eqref{e.mWp} to the order of $F>f$ as in Ref.~\cite{Schmaltz:2010ac}.

%%%%%%%%%%%%%%%%%%
\subsection{Fermion sector}
\label{s.top}
%%%%%%%%%%%%%%%%%%%%
In this framework, the SM fermions are elementary and their mass is obtained through Yukawa-like interactions with the composite Higgs as follows:
\beq
{\cal L}\supset \frac{\lambda_t}{\Lambda_t^{d-1}} Q_L t_R \op_H,	\label{Ltop}
\eeq
where $d$ is the scaling dimension of the Higgs operator \mbox{$\op_H\sim \hat\psi_3\tilde\psi$}, and $\Lambda_t$ is the scale where the above operator becomes strong, i.e., $\lambda_t\sim (4\pi)^2$. Below the conformal breaking scale, the Higgs operator confines with its VEV $\langle\op_H\rangle\sim \frac{\Lambda^d}{(4\pi)^2} \sin\theta$, generating mass for the top quark. The top quark mass can be expressed as:
\beq
m_t\equiv y_t v  \sim\frac{\lambda_t }{(4\pi)^2} \left(\frac{\Lambda}{\Lambda_t}\right)^{\!d-1}\, \Lambda\, \sin\theta, 	\label{e.mt}
\eeq
where $\Lambda\equiv 4\pi f$ is the scale of confinement/condensation, and $v =f\sin\theta$ is the SM Higgs VEV.
\footnote{An alternative possibility is the partial composite framework, where top-quark ($Q_L, t_R$) mixes linearly with the composite fermion operators $(\op_L,\op_R)$ as~\cite{Kaplan:1991dc,Vecchi:2015fma},
\[
{\cal L}\supset \frac{\lambda_L}{\Lambda^{d_L-5/2}} Q_L \op_L + \frac{\lambda_R}{\Lambda^{d_R-5/2}} t_R \op_R,
\]
where $d_{L,R}$ are the scaling dimensions of fermionic operator $\op_{L,R}$. The effective top Yukawa coupling is
\[
y_t\sim \frac{\lambda_L\lambda_R}{\Lambda^{d_L+d_R-5}}, 
\]
which is relevant for $d_{L,R}\lesssim5/2$~\cite{Vecchi:2015fma}.} 

In this framework, the flavor scale $\Lambda_t$ \eqref{Lam_t} is naturally decoupled from the strong dynamics responsible for the electroweak symmetry breaking for $\epsilon\equiv d-1\sim 1/{\rm few}$, e.g. for $\epsilon =1/3$ the flavor scale can be pushed up to $\op(10^3)\Lambda$, see also~\cite{Luty:2004ye}. For comparison, in a technicolor-like model, the scaling dimension is $d\sim3$ and therefore the flavor scale is $\op(5)\Lambda$ which would require $\Lambda\gg \op(1)\tev$ in order to avoid the flavor constraints in the absence of a flavor symmetry in the UV theory. Therefore, conformal UV completions of the little Higgs models or composite Higgs models naturally decouple the flavor scale from the electroweak breaking scale for scaling dimension $1\!<\!d\!<\!2$. 

Note that the top-quark mass operator of the form \eqref{Ltop} and other SM fermionic operators explicitly break the $SU(4)$ symmetry protecting the Higgs boson and other pNGB masses. We can write such operators in the UV such that they preserve the symmetries of the theory. For instance, each SM $SU(2)_L$ quark doublet, denoted as $Q_L=(u_L,d_L)^T$ with $u/d$ representing up/down-type quarks, originates from a fundamental representation of $SU(4)_L$. To complete the fundamental representation $SU(4)_L$, a second quark doublet, $Q^{\prime}_L=(u_L^\prime,d_L^\prime)$ is added, which transforms under $SU(2)_{L}^{\prime}$. Such that $\Psi_L\!=\!(Q_L,Q^\prime_L)^T$ transforms as a fundamental of $SU(4)_L$. The quark doublet $Q^{\prime}_L$ carries $U(1)_{Y}$ charges that are equivalent to the SM hypercharges of $Q_L$. Corresponding to each $Q^{\prime}_L$, there exists a field $Q^{\prime}_R=(U_R^\prime, D_R^\prime)^T$ that forms a vector-like pair. The SM $SU(2)_L$ singlet quark fields, $U_R, D_R$ and their partners $U_R^\prime, D_R^\prime$ only carry $U(1)_Y$ gauge quantum numbers. We can write them as $\Psi_R=(0,0,\lambda_u U_R, \lambda_d D_R)^T$ which transform as a fundamental of $SU(4)_R$. The quark Yukawa couplings can be derived from the following Lagrangian,
\begin{align}
\mathcal{L} &\supset \frac{1}{\Lambda_t^{d-1}}\bar\Psi_L (\hat \psi \psi) \Psi_R+\frac{\lambda^{\prime}}{4\pi} \Lambda \bar Q^{\prime}_L Q^{\prime }_R,\notag\\
&=  \frac{\Lambda}{(4\pi)^2} \left(\frac{\Lambda}{\Lambda_t}\right)^{d-1}\,\bar \Psi_L \Sigma \Psi_R+\lambda^{\prime} f \bar Q^{\prime}_L Q^{\prime }_R,\label{e.ytUV}
\end{align}
where $\Sigma$ transforms as $({\bf 4}, \bar {\bf 4})$ under the flavor symmetry ${SU(4)_L \times SU(4)_R}$. Above, $\lambda_{u,d}$ are strong couplings of order $(4\pi)^2$ through NDA estimates. Within this framework, one of the linear combinations of $U_R$ and $U^{\prime}_R \subset Q^{\prime}_R$ acquires a mass on the order of $f$ for $\lambda^\prime\sim1$, while the other linear combination corresponds to the conventional SM $SU(2)_L$ singlet up-type quark, $u_R$. A similar situation applies to the down-type quarks. The first term above leads to the generation of SM up-type and down-type quark masses. For instance, the top-quark mass is generated as given in equation \eqref{e.mt} with the top-Yukawa coupling,
\beq
y_t=\frac{\lambda_t}{4\pi} \left(\frac{\Lambda}{\Lambda_t}\right)^{d-1}\sim 4\pi  \left(\frac{\Lambda}{\Lambda_t}\right)^{d-1},
\eeq
for $\lambda_t\sim (4\pi)^2$. Note that the invariance of the first term under $SU(4)$ ensures the cancellation of one-loop quadratic divergences originating from the quark sector. Extending this approach to all quark generations and charged leptons follows a similar path.

%%%%%%%%%%%%%%%%%%
\subsection{Phenomenology}
\label{s.pheno}
%%%%%%%%%%%%%%%%%%%%
The low-energy phenomenology of this model closely resembles that of the `bestest' little Higgs model~\cite{Schmaltz:2010ac}, also discussed in~\cite{Ma:2015gra}. In this context, $H_1$ represents the SM Higgs doublet, while the second Higgs doublet $H_2$, as well as the scalar triplet $\Delta_1$ and singlet $\eta$, have substantial masses of the order $\mathcal{O}(1)$ TeV. Similarly, the heavy gauge boson partners $W^\prime$ and $Z^\prime$ also exhibit masses of approximately $\mathcal{O}(1)$ TeV. Additionally, the fermionic top partners have masses at the scale of $f$. These scalar, vector, and fermionic states, with masses around $\mathcal{O}(1)$ TeV, serve as promising targets for future LHC runs~\cite{Ma:2015gra}.

In this work, we do not conduct a detailed phenomenological study as it is highly model-dependent. Nevertheless, one distinctive characteristic of conformal little Higgs models, as opposed to other UV completions of little Higgs models, is the presence of broad resonances and continuum states associated with strong conformal dynamics at the scale $\Lambda$. For values of $f$ around $1$ TeV, which correspond to a tuning order of roughly $10\%$, the strongly coupled states with broad widths typically emerge at a scale of approximately $10$ TeV. Consequently, such states can be well-motivated targets for future colliders like FCC~\cite{FCC:2018byv,FCC:2018vvp}.

Here we briefly comment on potentially problematic low-energy phenomenological aspects that could differ from those of the `bestest' Little Higgs model~\cite{Schmaltz:2010ac}. First, we consider corrections to $Z\to b\bar b$ due to the first term in \eq{e.ytUV}, which generates the top-quark mass, as discussed in~\cite{Agashe:2006at}, see also~\cite{Galloway:2010bp,Agugliaro:2019wtf}. 
At the leading order, we have the following term in the effective Lagrangian,
\begin{align}
\mathcal{L}&\supset \frac{C_L}{(4\pi)^2} \left(\frac{\Lambda}{\Lambda_t}\right)^{\!2(d-1)}\mathrm{Tr}\Big[\bar \Psi_L \bar\sigma^\mu\Sigma^\dag \overleftrightarrow{D}_{\!\mu}\Sigma \Psi_L\Big],	\notag\\
&\supset \!\frac{C_L}{(4\pi)^2}\! \left(\frac{\Lambda}{\Lambda_t}\right)^{\!2(d-1)}\!\!\! \sin^2\!\theta\! \bigg[\! \frac{g_{\rm EW}}{\cos\theta_W}\!\Big(\!\bar t_L \gamma^\mu t_L - \bar b_L \gamma^\mu b_L\!\Big)Z_\mu 	\notag\\
&\hspace{2.7cm}+\frac{g_{\rm EW}}{\sqrt{2}}\Big(\bar t_L \gamma^\mu b_L W_\mu^+ +\text{h.c.}\Big)\bigg],
\end{align}
where $C_L\sim (4\pi)^2$ is the effective coupling in a strongly coupled theory and $\sin^2\!\theta=v^2/f^2$. The stringent constraint comes from the measurement of $Z\to b_L\bar b_L$~\cite{Gori:2015nqa}, which sets the following bound at the $2\sigma$ level,
\beq
\frac{\Delta g_{Zb_L\bar b_L}}{g_{Zb_L\bar b_L}^{\rm SM}}\sim \left(\frac{\Lambda}{\Lambda_t}\right)^{\!2(d-1)}\!\! \sin^2\!\theta\sim \left(\frac{y_t}{4\pi}\right)^{2} \frac{v^2}{f^2} \lesssim 1\%.
\eeq
We note that the above constraint is negligible and is always satisfied for $f>v$. Note that the extra suppression factor $(\Lambda/\Lambda_t)^{2(d-1)}\sim 1/(4\pi)^2$ is one of the core features of this framework, where the flavor scale $\Lambda_t$ is decoupled from the chiral (global) symmetry breaking scale $\Lambda=4\pi f$.
Similarly, for the top-quark contributions from the first term in \eq{e.ytUV}, the electroweak precision observable $T$-parameter is suppressed due to custodial symmetry. However, there are gauge and pNGB contributions similar to the original model~\cite{Schmaltz:2010ac}, as discussed in~\cite{Ma:2015gra}.
%%%%%%%%%%%%%%%%%%
\section{Comments and discussion}
\label{s.general}
%%%%%%%%%%%%%%%%%%%%
In the following comments, we argue that any known little Higgs model based on the coset $\G/\H$ where ${\cal F}\supset \G$ is weakly gauged can be made conformal little Higgs. The general idea is based on arguments presented in \cite{Thaler:2005kr,Cheung:2006ij,Cheng:2006ht}, see also \cite{Batra:2007iz}, where it is pointed out that any known little Higgs model based on the coset $\G/\H$ with weakly gauge subgroup ${\cal F}$ of $\G$ has the same low-energy physics as a two-site nonlinear sigma model with global symmetry $G^2$ spontaneously broken to $G$ and gauging the subgroups ${\cal F}\times\H$ in the limit that the gauge coupling of $\H$ is large. 
The latter case is realized provided $\G^2/\G$ breaking is due to a QCD-like confining dynamics.
Such a two-site model can be UV-completed with bifundamental fermions $\psi$'s with QCD-like confining gauge dynamics of $SU(N_c)$ or $Sp(2N_c)$ gauge symmetry as shown in the following Moose diagram
\beq
\vcenter{\hbox{\includegraphics[width=0.8\linewidth]{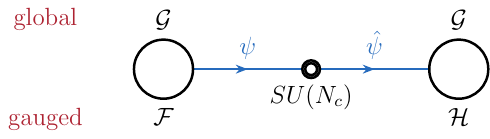}}}	\label{moose}
\eeq
A quick way to see this low-energy ``duality'' of a little Higgs model, based on global symmetry breaking coset $\G/\H$ with weakly gauged subgroup ${\cal F}$ (say theory-A) to that of a chiral breaking of $G^2\to G$ due to confining dynamics with gauged subgroup ${\cal F}\times H$ in the limit of large $\H$ gauge coupling (say theory-B), is counting the number of uneaten pNGBs in both theories. In theory A we have 
\beq
N_{\rm pNGB}^{\rm A}=\Big[\underbrace{N({\cal G})-N({\cal H})}_{\text{broken gen.}}\Big]-\underbrace{N({\cal F})}_{\text{eaten}},
\eeq
whereas, in theory B we have 
\beq
N_{\rm pNGB}^{\rm B}=\underbrace{N({\cal G})}_{\text{broken gen.}}\!\!-\Big[\underbrace{N({\cal H})+N({\cal F})}_{\text{eaten}}\Big].
\eeq 
Therefore the two theories have the same number of uneated pNGBs. Moreover, in the limit large $\H$ gauge coupling in theory-B, the corresponding gauge bosons are heavy and therefore can be integrated out similar to Hidden Local Symmetry~\cite{Bando:1987br}. Hence the two theories would give exactly the same low-energy dynamics with uneaten pNGBs and light gauge boson corresponding to ${\cal F}\subset \G$.

In order to UV complete the two-site little Higgs models (theory-B) with strongly coupled conformal dynamics one would need to extend the fermion sector of the confining gauge theory $SU(N_c)$ (or $Sp(2N_c)$) with additional bifundamental fermions $\chi$'s such that the theory lies in the conformal window with $N_f/N_c$ satisfying \eqref{CW}. In order to obtain a theory that undergoes spontaneous chiral symmetry breaking due to fermion condensate, it needs to exit the conformal fixed point and flow to the confining phase at scale $\Lambda$. Therefore, as argued above we would require a relevant deformation to the strongly coupled CFT in the form of \eqref{Ldef}. Such that at scale $\Lambda$ the additional massive fermionic d.o.f. $\chi$, responsible for conformal dynamics, can be integrated out. Hence the low-energy theory can be described as the above Moose diagram \eqref{moose}, where $\G^2$ symmetry is spontaneously broken to the diagonal subgroup $\G$ due to $\hat\psi\psi$ condensate of confining gauge theory $SU(N_c)$ (or $Sp(2N_c)$). The subgroups ${\cal F}\times \H$ are gauged and in the limit $\H$ gauge coupling is large, we get the low energy theory for a little Higgs model based on $\G/\H$ coset (theory A). 

Note that the moose \eqref{moose} would generally require that $\G^2/\G$ breaking is due to a QCD-like confining dynamics. Therefore this setup inherently requires $\G$ to be an $SU(N)$ flavor symmetry for $N$ fermion flavors in the fundamental representation with confining gauge theory $SU(N_c)$ (or $Sp(2N_c)$). In cases where the original little Higgs model has $\G$ as an $SO(N)$ flavor symmetry, one would require additional considerations. The feasibility of such an extension would, of course, depend on the specific model and its underlying dynamics. For instance, models based on $SO(N+1)$ with $N\in 2\mathbb{Z}$ can be recasted in $Sp(N)$, which can then be enhanced to $SU(N)$ symmetry. One can then explicitly break the $SU(N)/Sp(N)$ orthogonal symmetry which would make the corresponding pNGBs heavy. Then the low-energy theory would have $Sp(N)\sim SO(N+1)$ symmetry. An example of such a case with $SO(5)$ is noted above. Similarly for $SO(N)$ with $N\in 2\mathbb{Z}$ one can rely on their isomorphic or larger $SU$ groups.

It is relevant to make a comment here concerning the (large) hierarchy problem. Since all Higgs and Yukawa couplings dissolve, it implies that no scalars (pNGBs) propagate beyond the little Higgs scale $\Lambda$. Consequently, we are left with a chiral QCD-like theory, comprised of chiral fermions and non-abelian gauge fields. Symmetry breaking and condensation occur in the infrared (IR) following the logarithmic running of the gauge coupling. In other words, the symmetry breaking gives rise to the conformal anomaly (beta functions) below the little Higgs scale $\Lambda$, leading to the emergence of pNGBs and other low-energy degrees of freedom. As a result, the underlying theory resembles more of a chiral QCD scenario with no hierarchy problem and dimensional transmutation. Meanwhile, the little Higgs dynamics effectively addresses the little hierarchy between the electroweak scale $v$ and the conformal breaking scale $\Lambda$. 

%%%%%%%%%%%%%%%%%%
\section{Conclusions}
\label{s.conc}
%%%%%%%%%%%%%%%%%%%%
In this study, we have presented a UV completion for little Higgs models based on strongly coupled conformal dynamics, extending their validity to arbitrarily high energy scales. Our proposal suggests that any little Higgs model founded on the $\G/\H$ coset can achieve UV completion through robust conformal dynamics. The little Higgs mechanism ensures the absence of quadratic divergences for the SM Higgs doublet and other pNGB states up to the conformal breaking scale $\Lambda$. Beyond the conformal breaking scale, the theory displays conformal dynamics that persists up to an arbitrary high UV scale $M_{\rm UV}$, effectively addressing the gauge hierarchy problem up to this UV scale.

Furthermore, this framework naturally decouples flavor dynamics from the strong dynamics responsible for breaking the electroweak symmetry. With regards to the scaling dimension of the fermion condensate, which corresponds to the SM Higgs operator, in the range $1 < d < 2$, it has been demonstrated that the flavor scale can be exponentially separated from the confining scale $\Lambda$.

To illustrate the practicality of our concept, we have presented a concrete UV completion of the `bestest' little Higgs model based on the coset $SU(4)^2/SU(4)$. The low-energy phenomenology closely mirrors that of the canonical `bestest' little Higgs model. This model predicts the existence of a light SM Higgs doublet, a relatively heavy second Higgs doublet, a heavy $SU(2)_L$ triplet, and a singlet. The gauge partners of the SM gauge bosons have relatively heavy masses, typically of the order~$f$.

We have also discussed a potential UV completion for any known little Higgs model using strongly coupled conformal dynamics. The key features of this framework include the natural separation of the flavor problem and the possibility of detecting signatures of strong conformal dynamics at future colliders through broad resonances and continuum states at the conformal breaking scale $\Lambda$.

%%%%%%%%%%%%%%%%%%
\section*{Acknowledgements}
%%%%%%%%%%%%%%%%%%
We thank Zackaria Chacko and Jesse Thaler for their comments.
%%%%%%%%%%%%%%%%%%%%

%%%%%%%%%%%%%%%%
\bibliography{bib_CLH}{}
\bibliographystyle{aabib}
%%%%%%%%%%%%%%%%

\end{document}